# Atomic-scale patterning of hydrogen terminated Ge(001) by scanning tunneling microscopy

G Scappucci[1,2], G Capellini[3], W C T Lee[1] and M Y Simmons[1,2]

[1]School of Physics, University of New South Wales, Sydney, NSW 2052, Australia.

[2]Australian Research Council Centre of Excellence for Quantum Computer Technology, University of New South Wales, Sydney, NSW 2052, Australia.

[3]Dipartimento di Fisica, Università di Roma Tre, Via della Vasca Navale 84, 00146 Roma, Italy.

Email: giordano.scappucci@unsw.edu.au, michelle.simmons@unsw.edu.au

**Abstract**. In this paper we demonstrate atomic-scale lithography on hydrogen terminated Ge(001). The lithographic patterns were obtained by selectively desorbing hydrogen atoms from a H resist layer adsorbed on a clean, atomically flat Ge(001) surface with a scanning tunneling microscope tip operating in ultra-high vacuum. The influence of the tip-to-sample bias on the lithographic process have been investigated. Lithographic patterns with feature-sizes from 200 nm to 1.8 nm have been achieved by varying the tip-to-sample bias. These results open up the possibility of a scanning-probe lithography approach to the fabrication of future atomic-scale devices in germanium.

Atomic-scale patterning of H-terminated Ge(001) by STM

**1. Introduction**

Electronic device miniaturization to and beyond the 16 nm node requires technological progress other than standard scaling of silicon (Si) complementary metal-oxide-semiconductor (CMOS) technology [1]. This includes replacing Si in transistors with alternative high-mobility channel materials and developing technologies capable of miniaturizing devices at the atomic-scale, where atomically abrupt dopant profiles and interfaces are required [2].

For these reasons, there is a renewed interest in replacing the Si channel in transistors with germanium (Ge) due to its enhanced transport properties compared to Si and to the recent progress in the field of deposited high-κ dielectrics, developed within the Si CMOS platform, which has enabled gating of Ge-based transistors [3]. In particular, the lower effective masses of electrons and holes in Ge offer higher mobility (~3× for electrons and ~5× for holes, respectively) and higher saturation velocity (~7×), thus providing higher operating speed and larger driving current. Moreover, the lower bandgap (0.66 eV) compared to Si (1.12 eV) enables operation at lower voltages thereby minimizing power dissipation [4].

Important recent progress in Ge-based electronics has led to the demonstration of high quality p-channel transistors integrating Ge oxynitride [5] or hafnium oxide [6] as the gate dielectric; Ge/Si core/shell nanowire field-effect transistors [7] and single electron transistors [8]. However, a strategy for the fabrication of atomically precise electronic devices in Ge is yet to be developed.

Ultra-high vacuum (UHV) atomic-scale lithography in Si was demonstrated by using the highly confined electron beam from a scanning tunneling microscope (STM) tip to selectively desorb hydrogen (H) atoms from a H-terminated Si(001) surface [9, 10]. In this method, hydrogen atoms passivate the Si dangling bonds and act as a resist on the surface in analogy to that used in optical lithography. The possibility of selective doping by adsorption of $PH_3$ or $AsH_4$ on the exposed dangling bonds has led to proposals of integrating STM hydrogen lithography into atomic-scale semiconductor device fabrication [11]. Following this approach, silicon devices with nano- to atomic-scale phosphorus dopant profiles have been realized [12-15].

We recently demonstrated a low thermal budget technique for the fabrication of phosphorus delta-doped layers in Ge based on the adsorption and subsequent incorporation of $PH_3$ molecules onto a Ge(001) surface, followed by the encapsulation of the P atoms under a MBE-deposited epitaxial Ge layer [16]. The ability to electrically dope the Ge(001) surface with a gas phase dopant source is analogous to the process developed for atomic-scale silicon devices [17]. In this paper we show that STM hydrogen lithography is a viable approach for controllably patterning nano- to atomic-scale features on a H-terminated Ge(001) surface. STM hydrogen lithography on Ge(001) is a key milestone that, combined with the phosphorus delta-doping technique, opens up the possibility of a UHV-STM approach to the fabrication of atomic-scale devices in Ge. While the adsorption of hydrogen and STM lithography on Si(001) have been studied extensively in the past, only a few STM studies of hydrogen on Ge(001) have been reported [18-21]. Earlier work [18] has shown the formation of a well ordered 2×1:H monohydride phase on Ge(001) with 1 monolayer of hydrogen coverage, very similar to the monohydride terminated Si(001). Obtaining a uniform low-defected H-terminated surface, however, depends critically on the hydrogen dose. Exposure to H at high doses can result in etching of the surface caused by the atomic H generating Ge atom/dimer vacancies that subsequently evolve in line defects along the Ge dimer rows [19]. Hydrogen coverages of less than one monolayer, instead, were used recently in Ref [20] and [21] to study, respectively, the properties of an isolated Ge dangling bond formed by the chemisorption of a single H atom on a Ge dimer [20] and the variety of adsorption structures of $H_2$ and H on Ge(001) [21].

Modification of H-terminated Ge surfaces with scanning probes has been limited so far to micron-scale features on Ge(001), patterned in an ambient nitrogen atmosphere with no atomic resolution STM image [22], or isolated hydrogen atoms desorbed from Ge(111) [23]. We exploit the imaging capability of the UHV-STM to optimize the initial Ge(001) surface preparation and hydrogen termination procedure and obtain an atomically-flat low defect monohydride saturated Ge(001) surface. On such surfaces we investigate the use of the STM tip to stimulate the desorption of hydrogen atoms and show how using different lithographic parameters we achieve lithography with feature-sizes from 200 nm to 1.8 nm. The possibility of creating, within these areas, ultra-sharp *n*-type doping profiles for use as atomically abrupt source/drain regions for high-mobility atomically-precise Ge transistors is particularly interesting for ultra-scaled CMOS technology.

**2. Experimental**

All experiments were carried out in an ultra-high vacuum STM system with a base pressure $< 5 \times 10^{-11}$ mbar. Ge(001) samples $2.5 \times 10$ mm$^2$ in size were cleaved from a Sb doped Ge(001) 4 inch wafer (resistivity of 1-10 Ωcm). Homogeneous substrate heating was achieved by flowing a DC current $I_{DC}$ directly through the sample and monitoring the temperature with an infrared pyrometer. For temperatures below the detection limit of the infrared pyrometer ($T < 250°C$), we took care to measure in detail, for each sample, the $T$ vs. $I_{DC}$ calibration curve for $T > 250 °C$ and then extrapolate from $T = 250 °C$ to room temperature. The extrapolation procedure was checked using a thermocouple mounted on the sample holder. For the temperature values reported in figure 2 a measurement uncertainty of ±10 °C should be taken in account.

For initial preparation of a clean Ge(001) surface we used a two-step method of *ex-situ* chemical passivation followed by an *in-situ* heating procedure. A GeO$_x$ passivation layer was chemically grown *ex-situ* by a wet treatment consisting of a HCl:H$_2$O 36:100 bath and subsequent H$_2$O$_2$:H$_2$O 7:100 bath to strip/reform a GeO$_2$ passivation layer. The samples were then outgassed *in-situ* at 230 °C for ~ 1 h, flash-annealed at 760 °C for 60 s to remove the GeO$_2$ and slowly cooled ( at ~ 2 °C s$^{-1}$) from 600 °C to room temperature to obtain an ordered reconstructed surface. During the *in-situ* thermal treatment the pressure in the vacuum chamber was below $3 \times 10^{-9}$ mbar.

For H-termination of the surface we used an atomic hydrogen source consisting of a tungsten filament heated to 1400 °C, located ~ 15 cm in front of the surface and thermally shielded by means of a water heat shroud. After the initial flash-anneal at 750 °C, the sample was slowly cooled from 600 °C to the targeted process temperature. Different doses were obtained by regulating through a precision leak valve the hydrogen pressure in the chamber to a value of $5\times10^{-7}$ mbar and by changing the exposure time. The doses are given in Langmuirs of H$_2$ because, with this setup, the measured chamber pressure increment is mainly due to the presence of H$_2$.. After H-termination the samples were cooled to room temperature and transferred to the chamber for STM characterization.

Filled state atomic resolution STM measurements were performed at room temperature with typical sample bias $V_{sample}$ ~ -1.3 to -3 V and tunneling current $I$ in the 0.2 – 0.5 nA range.

## 3. Results and discussion

The preparation of a low-defected starting Ge(001) surface with minimal roughness is the first critical step for the subsequent formation of a uniform hydrogen resist layer suitable for investigation of STM lithography on different length scales.

In figure 1(a)-(b) we report typical STM images of the as prepared Ge(001) surfaces on two different length scales. Investigations of the topography on the 150×150 nm$^2$ scale [figure 1(a)] reveal the presence of uneven mono-atomic terraces having a broad width distribution with irregular edges, owing to the tendency of the [010]-bounded terraces to form kinks during the high temperature preparation process, in agreement with previously published results [24]. From the analysis of the STM image we calculate an average height difference from peak to valley of ~ 1 nm (which corresponds to ~7 atomic steps) and a root mean square surface roughness ~ 0.95 Å. We observe randomly distributed small mounds that appear as bright features in figure 1(a), with a typical area of ~ 3×3 nm$^2$ and an apparent height of ~5 Å, as reported previously for Ge(001) surfaces prepared with different methods [25, 26]. The exact origin of these features will be the subject of a future detailed STS study.

The atomic resolution STM image of the Ge(001) surface in figure 1(b) shows the typical p(2×1) dimer rows ~8 Å apart visible on a terrace ~ 1.4 Å high. The defects on the surface are mainly missing dimer rows with a concentration of ~ 2 %. We also observe the expected higher order p(2×2) and c(4×2) reconstruction due to in-phase or out-of-phase buckling of surface dimers on adjacent dimer rows with respect to the surface plane. In buckled (asymmetric) dimers the bond direction tilts out of the surface plane. This results in an opening of a gap between occupied and unoccupied surface states associated with an effective charge transfer (~0.1e) from the down atom to the upper atom that lowers the energy of the dimer thereby stabilizing the surface [24, 27].

The low-defected surface shown in figure 1 is now suitable for the next step of the process, i.e. the adsorption of a monolayer of hydrogen to act as a resist layer for STM lithography. For the purpose of creating a uniform H-terminated Ge(001) surface we are interested in obtaining a full surface coverage with the monohydride phase, where each H atom is bound to a Ge atom at the end of a dimer forming a 2×1 reconstruction. Note that this is a true symmetric 2×1 reconstruction where the H atoms saturate the Ge dangling bonds preventing charge transfer and buckling in the dimer so that higher order surface reconstructions are not observed. Previous studies have shown that a monohydride-saturated Ge(001) surface can be achieved by exposure to high hydrogen dose (~1000 L) to obtain full coverage while keeping the process temperature in the 100-200°C range to avoid the formation of higher hydrides [18]. Exposure at high doses, however, can degrade the surface morphology due to etching of the Ge surface caused by the atomic H impinging the surface [19].

Figure 2 summarizes the results of a series of experiments (H1-H5) in which we varied the process temperature and hydrogen dose, indicated in the inset of figure 2, to obtain a low defect density monohydride surface. For each experiment we plot the percentage of surface defects calculated as the fraction of surface area occupied by defects relative to the whole surface area scanned by STM (typically 200×200 nm$^2$). Close to each point in figure 2 we show a corresponding high resolution 25×25 nm$^2$ STM image of the surface after termination. The defects on the surface are single atom vacancies (AV), dimer vacancies (DV) and 1D strings of dimer vacancies (line defects, LD), which run along the dimer rows and appear as dark depressions in filled states STM images.

At a dose of 450 L and process temperature of ~ 100 °C (H1) defects cover ~ 15 % of the surface, which shows a monohydride 2×1 reconstruction with line defects aligned along the Ge dimer rows due to significant etching caused by atomic hydrogen. Under these dosing conditions the line defects extend up to ~ 4 nm along the Ge dimer rows. Previous studies have explained the mechanism giving rise to these line defects: Ge dimers on the surface form instable GeH$_2$ dihydrides pairs that - via successive H addition - desorb and generate dimer vacancies. The adjacent dimers are destabilized by the tensile strain that results from this desorption and strings of dimer vacancies then grow parallel to the dimer row by successive removal of neighboring dimers [19]. Line defects aligned along dimers caused by atomic hydrogen induced etching have also been observed previously on Si(001) [28]. We also observe bright protrusions, randomly distributed over the surface with an apparent height of ~ 0.8 Å that we attribute to the presence of single dangling bonds (DB) due to incomplete termination of the surface. Decreasing the dose to 135 L (H2) reduces the damage from the atomic H to < 10% of the surface, but with significantly more dangling bonds. Increasing the process temperature to 140 °C (H3) and 180 °C (H4) allows higher adatom mobility on the surface that mitigates the damage associated with the etching action of the atomic hydrogen. As a consequence the line defects are shorter (typically < 1 nm) than those observed in experiment H1. The surface percentage occupied by defects (Exp. H4) is now comparable with that obtained on the clean starting Ge(001) surface. A further increase in process temperature would trigger thermal desorption of H atoms and was therefore avoided [29]. Finally we reduced the hydrogen dose to 65 L (H5) and obtained, as seen in the STM image, a high quality 2×1 reconstructed monohydride surface. The defects are mainly dimer row vacancies that occupy less than 1% of the surface area, better than that obtained on the clean Ge(001) surface, suggesting that under appropriate dose/temperature parameters, exposure to atomic hydrogen can result in surface self-healing. The dangling bond density is also low, less than 1% of a monolayer. Such a low-defect density surface provides an ideal resist layer for STM lithography and all subsequent experiments described were performed on monohydride Ge surfaces prepared with the H-termination procedure of experiment H5.

To selectively desorb hydrogen atoms from the pristine resist layer, we positioned the tip in tunneling condition over the monohydride Ge(001) surface, changed the sample bias and tunneling current to predetermined desorption parameters and moved the tip at a constant speed in straight lines between points to write the desired lithographic pattern. After that we set the STM feedback loop back to imaging conditions and scanned the tip over the patterned area. For lithography we used a tunneling current of 3 nA and tip speed of 100 nm/s and investigated the effect of the sample bias on the desorption linewidth by writing a line pattern at sample biases of +6.5 V, +5 V and +4.5 V.

The alphabetical letter "G" [figure 3(a)] was written at sample bias of +6.5 V moving the tip along the 350-nm-long line pattern indicated with a dotted line. The desorbed areas are brighter than the surrounding saturated dimers due to the additional contributions to the tunneling current from the surface states of the exposed Ge dimers. On this 145×130 nm$^2$ large-scale image the contrast between patterned/unpatterned areas is weak, because the patterned region extends over several terraces of the monohydride surface. The larger bright features in figure 3(a) a few-nm-wide were present on the surface before hydrogenation whereas the smaller bright spots are caused by additional spurious desorption of the hydrogen resist during patterning due to stray electrons from the STM tip. A close-up image of the desorbed pattern over a single terrace is shown in figure 3(b), with a clear contrast between desorbed region and monohydride surface. Atomic resolution STM images within the desorbed areas indicate the restoring of a clean atomically flat Ge(001) surface, characterized, as a fingerprint, by the reappearance of extended c(4×2) domains perfectly reconstructed [figure 3(c)]. 2D Fourier transforms of the STM images of the pristine background resist and of the desorbed areas [figure 3(d) and (e) respectively] highlight the different surface reconstructions for the monohydride and the clean Ge(001) surface, the latter showing extra peaks due to the domains of the higher symmetry c(4×2) unit cell.

Figure 3(f) shows the letter "G" written at a sample bias of +5 V on a 65×65 nm$^2$ area of the surface extending, for almost its entirety, on a single mono-atomic terrace. Decreasing the sample bias sharpens the pattern boundaries and limits spurious desorption events. Finally, the letter "C" in figure 3(g) was written by further decreasing the sample bias to +4.5 V. As an inset in figure 3(g) we report a high resolution image of a line pattern desorbed under identical conditions. At a desorption sample bias of +4.5 V we observe the highest lithographic pattern quality. Further reduction of the sample bias leads to incomplete desorption and interrupted lithographic lines.

The average height profiles measured along the line patterns desorbed at different sample bias $V_{sample}$ are plotted on the same graph in figure 3(h) vs. the lateral position, offset for clarity for each trace. The letter "G" desorbed at $V_{sample}$ = +6.5 V has a broad average height profile [triangles] and we estimate an average linewidth $w \sim 8.4$ nm from the width of the central flat region showing an apparent average height difference between H-terminated and H-removed areas of ~ 0.8 Å. Note: the average height difference between H-terminated and H-removed areas is calculated with the height in the H-removed areas averaged along the desorption path over symmetric (2×1) and buckled c(4×2) Ge dimers. For the letter "G" desorbed at $V_{sample}$ = +5 V the average height profile [circles] shows a sharp peak. We measure an average linewidth $w = 4.7 \pm 0.3$ nm from the full-width at peak maximum of the Gaussian curve fitted to the experimental points. Finally, for the letter "C", reduction of $V_{sample}$ to +4.5 V results in a narrow peaked average height profile [squares]. With the same procedure described above we measure an average linewidth $w = 1.80 \pm 0.07$ nm, which corresponds to the spacing of 2-3 Ge dimer rows. The upper-right inset graph in figure 3(h) summarizes the average linewidth $w$ vs. sample bias dependence for the three lithographic patterns discussed above. At fixed tunneling current and tip speed we clearly observe that increasing the sample to tip bias increases the pattern linewidth. Given a set of desorption parameters, however, the exact linewidth is dependent on the particular shape of the tungsten tip being used, as has been reported for STM H-lithography on Si(001) [30]. For silicon it is well known that the tunneling current affects the lithographic linewidth [9, 10, 30]. At a fixed low sample bias, continuous lithographic lines 1nm-wide were obtained in the H:Si(001) system by increasing the tunneling current, resulting in a closer tip to sample spacing [9]. The closer tip to sample spacing leads to a reduced electron beam divergence and, ultimately, to an improved resolution. Future studies will investigate the influence of the tunneling current on pattern linewidth in the H:Ge(001) system with the goal of improving the patterning resolution as for the H:Si(001) system.

At a bias of +4.5 V we achieve a minimum linewidth of 1.8 nm as observed for the line pattern "C" in figure 3(g). At this bias the voltage on the surface is lower than the work function for Ge (~ 5 eV [31]) and hydrogen desorption occurs when the Ge-H bond is broken due to the inelastic interaction between electrons tunneling from the tip and adsorbed hydrogen atoms in agreement with a low-energy electron stimulated desorption mechanism. Due to the low Ge-H binding energy (3.0 eV [32]), the electronic excitation of a single electron in the $\sigma^*$ (Ge-H) antibonding unoccupied orbital at 3.5 V above the Fermi level [33] is sufficient - in the low-bias tunneling regime - to produce the Ge-H bond breaking, as indicated by previous studies of single hydrogen atom desorption from Ge(111) surfaces [23]. This is different in the Si(001)-H system, where the desorption mechanism in the low-bias regime involves multiple-vibrational excitations by tunneling electrons due to the higher Si-H binding energy (3.5 eV [32]) [10].

In the low-bias regime high spatial lateral resolution is achieved because of the close distance between tip and sample and the virtual absence of scattering effects that are the resolution limiting factor of conventional electron beam lithography. As the sample bias increases above +5 V [letter "G" in figure 3(a) and (f)] we observe a pattern linewidth increase to ≥ 5 nm due to the larger sample/tip distance, which spreads the electron beam to a larger area, and because we are above the work function of the surface, the STM is operating in field emission regime. In this regime electrons propagate in a high field vacuum gap and hit the surface generating secondary/reflected electrons which are likely to return to the sample and cause additional desorption of the hydrogen resist. STM desorption in the field emission regime with linewidths limited to ~ 5 nm have been previously reported for H-terminated Si(001) [10].

The flexibility of the STM lithography is demonstrated in figure 3(i) and (j) where regions of exposed Ge(001) of different sizes were obtained by controlling differently the tip movements on the H resist. The 200×200 nm$^2$ large square in figure 3(i) was desorbed by rastering the tip maintaining constant desorption parameters ($V_{sample}$= +7.5 V, $I$ = 3 nA, linear speed of 100 nm/s). The small ~ 5×5 nm$^2$ dot [figure 3(j)] containing ~ 80 Ge dangling bonds was obtained by pulsing the tip – without moving it – to desorption parameters ($V_{sample}$= +5 V, $I$ = 3nA) and then back to imaging conditions.

The ability to pattern features over these two length scales using STM lithography has proven crucial for the fabrication of nano- to atomic-scale devices in silicon: large square regions selectively doped with phosphorus have been consecutively patterned on H-terminated Si(001) to aid the formation of ohmic contacts to atomic-scale donor quantum dots containing < 10 electrons [34]. As a comparison, by electrically doping the Ge dot pattern in figure 3(j) with phosphine gas we would obtain a donor quantum dot in Ge with ~ 10 electrons, estimated from the dot area and the electron density $5\times10^{13}$ cm$^{-2}$ achieved in the Ge:P delta-layers [16].

## 4. Conclusions

In summary, we have demonstrated the use of a scanning tunneling microscope to pattern nano- to atomic-scale features on a H-terminated Ge(001) surface. To this end, we have shown the careful optimization of both the initial Ge(001) surface preparation and hydrogen termination procedure to obtain an atomically flat and low defect passivated surface suitable for atomic-scale lithography. Using different lithographic patterning parameters we have achieved a minimum linewidth of 1.8 nm. Within the nanoscale areas patterned by STM we demonstrate the restoration of a clean atomically flat Ge(001) that opens up the possibility of using the reactive dangling bonds surrounded by the relatively inert hydrogen resist for selective chemistry of the Ge(001) surface. By combining the STM H-lithography technique demonstrated here with the phosphorus delta-doping we have recently demonstrated [16] we intend to achieve selective doping of Ge with ultra-sharp profiles, a key milestone towards the fabrication of nano- to atomic-scale electronic devices in Ge.

**Acknowledgments**

GS acknowledges support from UNSW under the 2009 Early Career Research/Science Faculty Research Grant scheme. GS acknowledges J. Miwa for fruitful discussions. GC is thankful to UNSW for a Visiting Professor Fellowship. MYS acknowledges an Australian Government Federation Fellowship and a Linkage grant with Zyvex Laboratories.

Atomic-scale patterning of H-terminated Ge(001) by STM

**Figure 1**

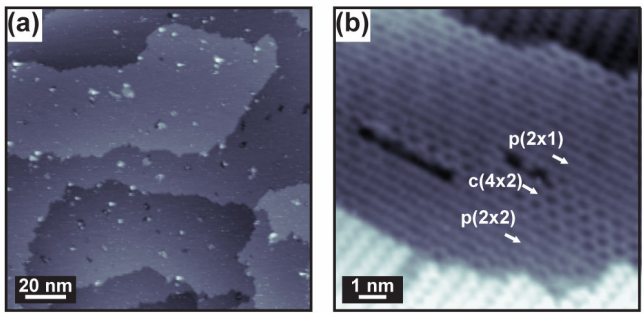

Figure 1. (Color online) STM images of the as prepared Ge(001) surface at different length scales. Images in (a) and (b) are 150×150 nm$^2$ and 11×11 nm$^2$ respectively. The images were acquired with a sample bias of -2.4V and a tunneling current of 0.47 nA.

**Figure 2**

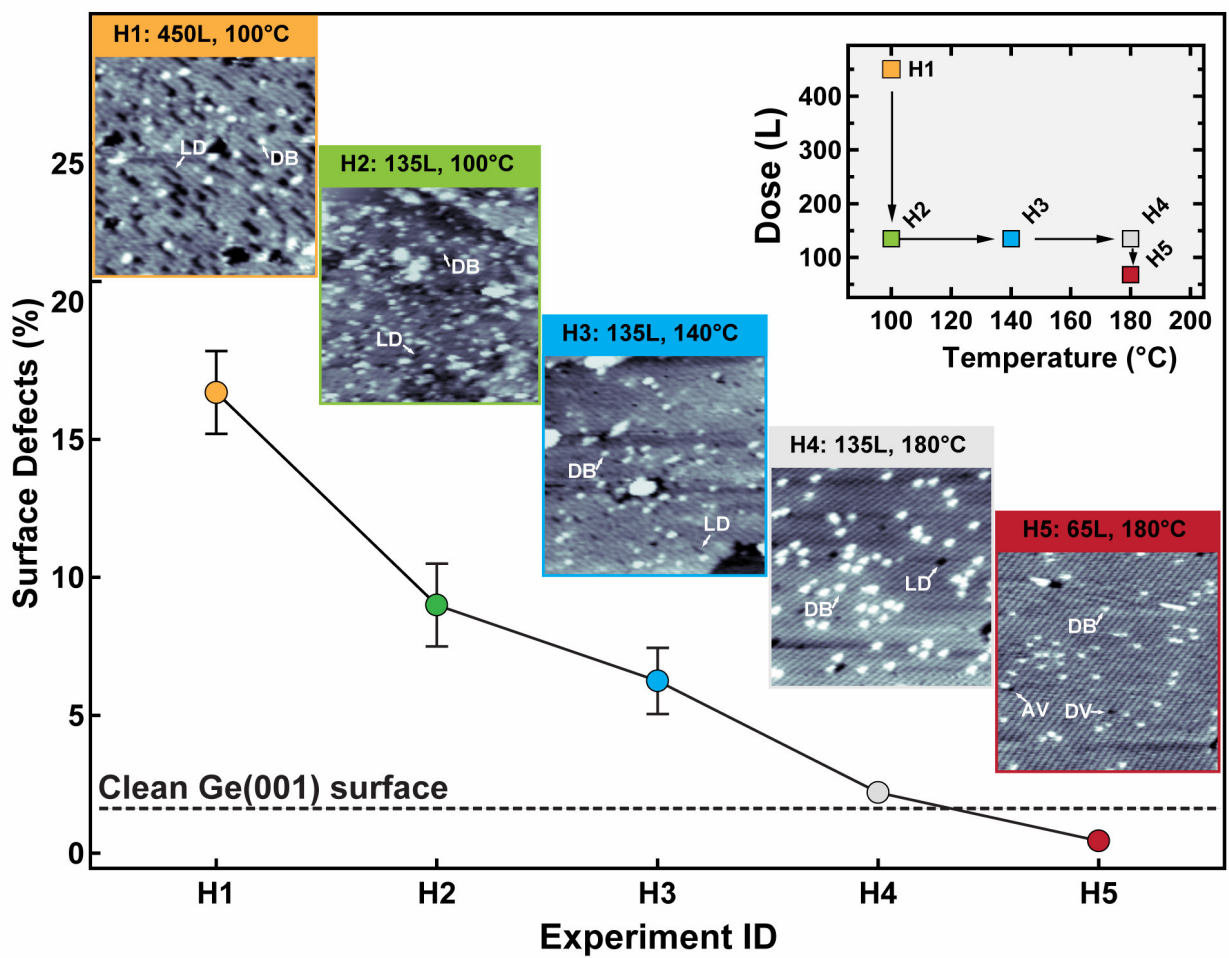

Figure 2. (Color online) A series of results showing the percentage of surface defects (see text for details) and corresponding 25×25 nm$^2$ STM images of the Ge(001) surface after 5 different experiments with different hydrogen ($H_2$) doses and process temperature (labeled H1-H5), shown in the upper-right inset. The images were acquired at the following sample bias/tunneling current: -2.2 V/0.32 nA (H1), -2.5 V/0.37 nA (H2), -2.2 V/0.48 nA (H3), -2.4 V/0.47 nA (H4), -2.4 V/0.41 nA (H5).

Atomic-scale patterning of H-terminated Ge(001) by STM

**Figure 3**

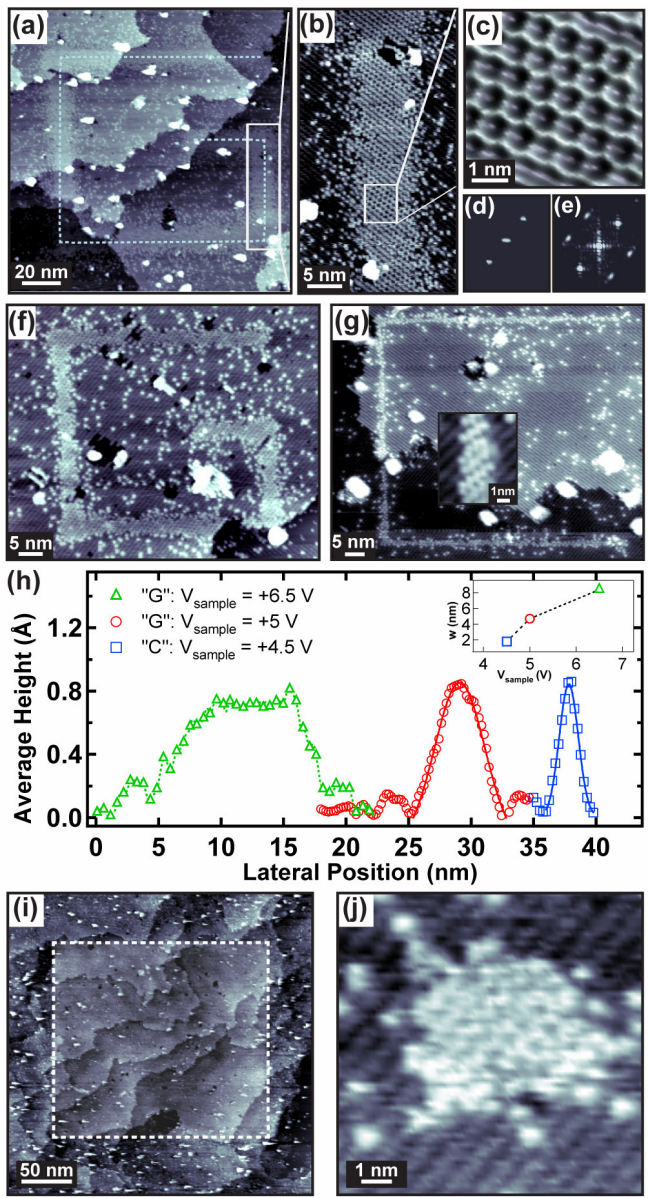

Figure 3. (Color online) Atomic-scale lithographic patterns on H-terminated Ge(001). (a) Large-scale and (b), (c) atomic-scale resolution images of the letter "G" desorbed at a STM tip to sample bias of +6.5V with 2D Fourier transforms of (d) the H-resist and (e) desorbed area showing the Ge(001) surface reconstruction. (f), (g) Letters "G" and "C", with high resolution inset, written reducing the sample bias to +5 V and +4.5 V respectively. (h) Average height profiles of the above lithographic patterns along the desorbed lines with (inset) average linewidth $w$ vs. sample bias dependence. (i) Square and (j) dot lithography done by rastering or pulsing respectively the tip at constant desorption parameters (see text for details). The images were acquired at the following sample bias/tunneling current: -2.4 V/0.46 nA [(a)-(c), (f), (j)]; -2.2 V/0.41 nA (g); -1.3 V/ 0.3 nA [inset in (g)]; -2.7 V/0.4 nA (i).